\documentclass[12pt]{article}
\usepackage{pdproc,epsfig} 

\def\be{\begin{equation}}
\def\ee{\end{equation}}
\def\beq{\begin{equation}}
\def\eeq{\end{equation}}
\def\bea{\begin{eqnarray}}
\def\eea{\end{eqnarray}}
 
\begin{document}

\renewcommand{\thefootnote}{\alph{footnote}}
  
\title{Neutrino Mass Models:\\ Impact of non-zero reactor angle}

\author{Stephen~F.~King}

\address{
School of Physics and Astronomy,
University of Southampton,\\
Southampton, SO17 1BJ, U.K. \\
 {\rm E-mail: king@soton.ac.uk}}

  \abstract{In this talk neutrino mass models are reviewed and the impact of a non-zero reactor angle
  and other deviations from tri-bimaximal mixing are discussed. We propose some benchmark models, where 
the only way to discriminate between them
is by high precision neutrino oscillation experiments.}
   
\normalsize\baselineskip=15pt

\section{Introduction}

\begin{figure}[htb]
\centering
\includegraphics[width=0.86\textwidth]{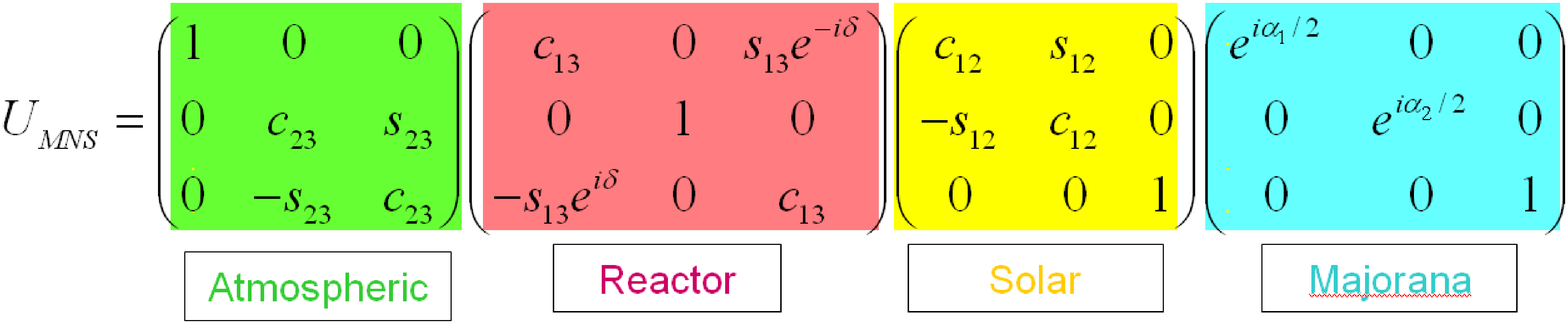}
\vspace*{-4mm}
    \caption{The lepton mixing matrix with phases factorizes
into a matrix product of four matrices,
associated with the physics of Atmospheric neutrino oscillations,
Reactor neutrino oscillations, Solar neutrino oscillations
and a Majorana phase matrix.} \label{MNS2}
\vspace*{-2mm}
\end{figure}

 In the three active neutrino paradigm, the lepton mixing matrix can be parameterised as in Fig.\ref{MNS2}
in terms of three angles $\theta_{ij}$, one oscillation phase $\delta$ and (if neutrinos are Majorana particles)
two Majorana phases $\alpha_i$.

%%%%%%%%%%%%%%%%%%%%%%%%%%%%%%%%%%%%%%%%%%%%%%%%%%%%%%%%%%%%%%%%%%%%%%
%%%%%%%%%%%%%%%%%%%%%%%%%%%%%%%%%%%%%%%%%%%%%%%%%%%%%%%%%%%%%%%%%%%%%%

\begin{figure}[h]
\begin{minipage}{18pc}
\includegraphics[width=18pc]{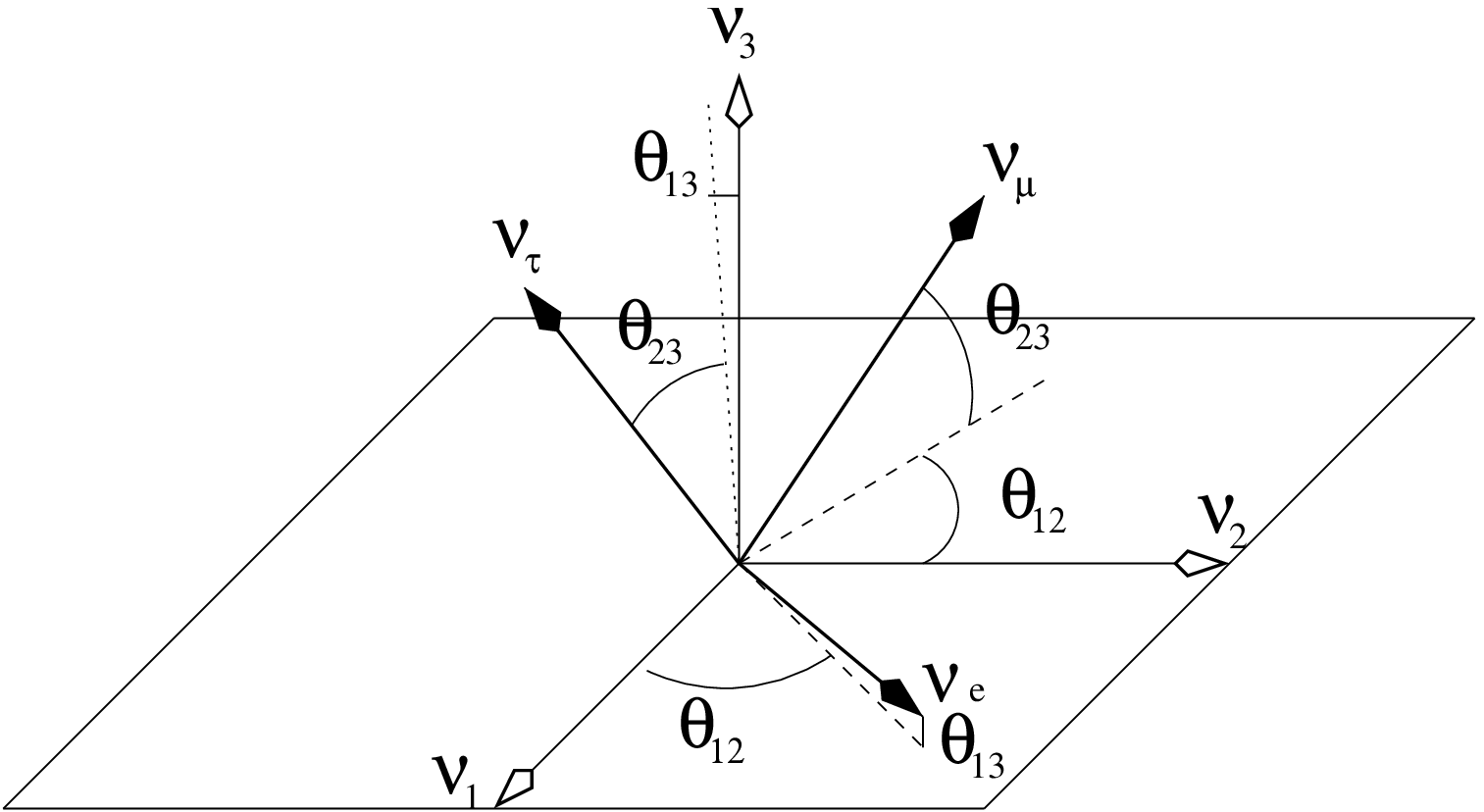}
\caption{\label{angles}The relation between
the neutrino weak eigenstates $\nu_e$, $\nu_\mu$, and $\nu_\tau$ and
the neutrino mass eigenstates $\nu_1$, $\nu_2$, and $\nu_3$
in terms of the three mixing angles $\theta_{12}$,
$\theta_{13}$, $\theta_{23}$.
Ignoring phases, these are just the Euler angles
respresenting the rotation of one orthogonal basis
into another.}
\end{minipage}\hspace{2pc}%
\begin{minipage}{14pc}
\includegraphics[width=14pc]{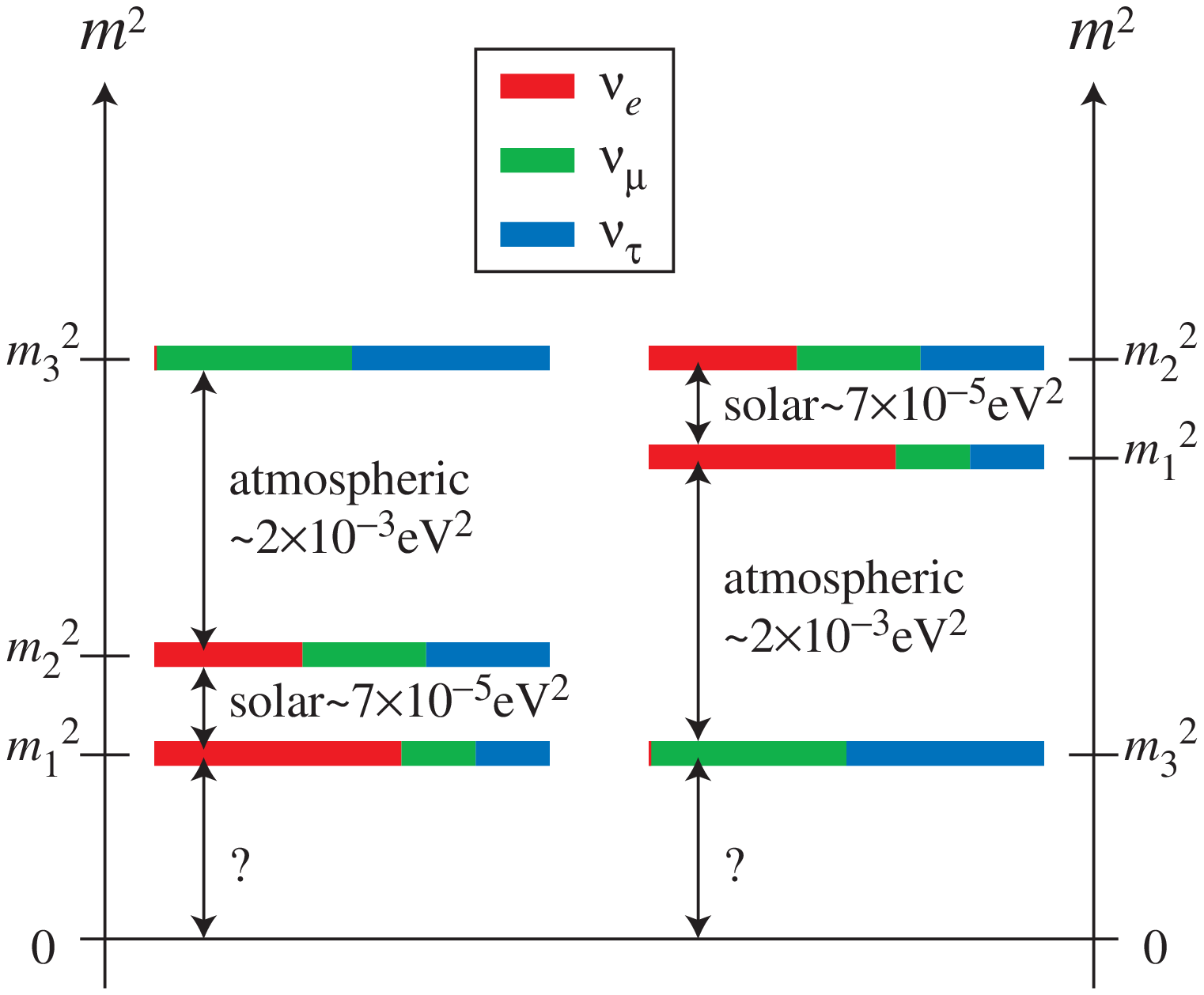}
\caption{\label{mass}
Alternative neutrino mass
patterns that are consistent with neutrino
oscillation explanations of the atmospheric and solar data.
The pattern on the left (right) is called the normal (inverted)
pattern. The coloured bands represent the probability
of finding a particular weak eigenstate
$\nu_e$, $\nu_\mu$, and $\nu_\tau$
in a particular mass eigenstate.}
\end{minipage} 
\end{figure}

\begin{figure}[h]
\begin{minipage}{12pc}
\includegraphics[width=12pc]{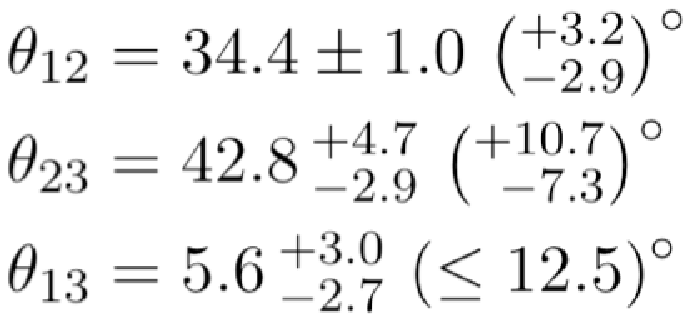}
\caption{\label{angle}  The best fit lepton mixing angles with 1$\sigma$ error
    (3$\sigma$ error).    }
\end{minipage}\hspace{2pc}%
\begin{minipage}{20pc}
\includegraphics[width=20pc]{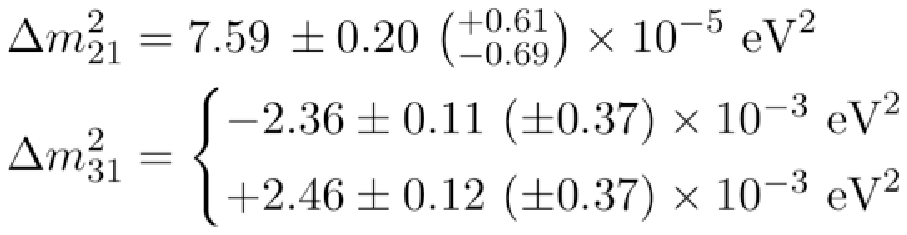}
\caption{\label{masses}
The best fit neutrino mass squared differences with 1$\sigma$ error
    (and 3$\sigma$ error).
    %from \cite{GonzalezGarcia:2010er}.    
    }
\end{minipage} 
\end{figure}

Ignoring the phases, the lepton mixing angles can be visualised as the Euler angles in Fig.\ref{angles}.
The mass squared ordering is not yet determined uniquely for the atmospheric mass squared splitting,
but the solar neutrino data requires $m_2^2 > m_1^2$, as shown in Fig.\ref{mass}.
The absolute scale of neutrino masses is not
fixed by oscillation data
and the lightest neutrino mass may vary from about $0.0-0.2$ eV where the upper limit
comes from cosmology. The current best fit values for the lepton angles and neutrino mass squared differences
are given in Figs.\ref{angle},\ref{masses}. Note that the reactor angle $\theta_{13}$ is not currently measured but its value is only inferred. 
%The prospects for its future determination have been projected as in Fig.\ref{future}.

%\begin{figure}[h]
%\includegraphics[width=14pc]{future.eps}\hspace{2pc}%
%\begin{minipage}[b]{14pc}\caption{\label{future}
%The future $\sin^22\theta_{13}$ sensitivity limit (normal hierarchy, 90\% CL)
   % from \cite{Huber:2009cw}.  }
%\end{minipage}
%\end{figure}

\section{Why go beyond the Standard Model?}

It has been one of the long standing goals of theories of particle
physics beyond the Standard Model (SM) to predict quark and lepton
masses and mixings. With the discovery of neutrino mass and
mixing, this quest has received a massive impetus. Indeed, perhaps
the greatest advance in particle physics over the past decade has
been the discovery of neutrino mass and mixing involving two large
mixing angles commonly known as the atmospheric angle
$\theta_{23}$ and the solar angle $\theta_{12}$, while the
remaining mixing angle $\theta_{13}$, although unmeasured, is
constrained to be relatively small. The largeness of the two large
lepton mixing angles contrasts sharply with the smallness of the
quark mixing angles, and this observation, together with the
smallness of neutrino masses, provides new and tantalizing clues
in the search for the origin of quark and lepton flavour. However,
before trying to address such questions, it is worth recalling why
neutrino mass forces us to go beyond the SM.

Neutrino mass is zero in the SM for three independent reasons:
\begin{enumerate}
\item There are no right-handed neutrinos $\nu_R$. \item There are
only Higgs doublets of $SU(2)_L$. \item There are only
renormalizable terms.
\end{enumerate}
In the SM these conditions all apply and so neutrinos are massless
with $\nu_e$, $\nu_{\mu}$, $\nu_{\tau}$ distinguished by separate
lepton numbers $L_e$, $L_{\mu}$, $L_{\tau}$. Neutrinos and
antineutrinos are distinguished by total conserved lepton number
$L=L_e+L_{\mu}+L_{\tau}$. To generate neutrino mass we must relax
one or more of these conditions. For example, by adding
right-handed neutrinos the Higgs mechanism of the Standard Model
can give neutrinos the same type of mass as the electron mass or
other charged lepton and quark masses. It is clear that the {\it
status quo} of staying within the SM, as it is usually defined, is
not an option, but in what direction should we go?

 \section{Tri-bimaximal mixing}
It is a striking fact that current data on lepton mixing is
(approximately) consistent with the so-called tri-bimaximal (TBM)
mixing pattern \cite{Harrison:2002er},
\begin{equation}
\label{TBM}
U_{TB}= \left(\begin{array}{ccc} \sqrt{\frac{2}{3}}& \frac{1}{\sqrt{3}}&0\\
-\frac{1}{\sqrt{6}}&\frac{1}{\sqrt{3}}&\frac{1}{\sqrt{2}}\\
\frac{1}{\sqrt{6}}&-\frac{1}{\sqrt{3}}&\frac{1}{\sqrt{2}}
\end{array} \right) P_{Maj},
\end{equation}
where $P_{Maj}$ is the diagonal phase matrix involving the two
observable Majorana phases. However in realistic models tri-bimaximal mixing cannot be exact
and deviations from tri-bimaximal mixing can be parametrized by three parameters $r,s,a$
defined as \cite{King:2007pr}:
\begin{eqnarray}
&& \sin \theta_{13} =   \frac{r}{\sqrt{2}}, \ \ \sin \theta_{12} =
\frac{1}{\sqrt{3}}(1+s), \ \  
\sin \theta_{23}  =  \frac{1}{\sqrt{2}}(1+a). \label{rsa}
\end{eqnarray}
The global fits of the conventional mixing angles \cite{GonzalezGarcia:2010er}
can be translated into the
$1\sigma$ ranges:
\begin{eqnarray}
&& 0.07<r<0.21,\ \ -0.05<s<0.003, \ \  -0.09<a<0.04.
\end{eqnarray}
Tri-bimaximal mixing corresponds to $ \theta_{12} = 35^o$, 
$\theta_{23}  =  45^o$ and $\theta_{13}  =  0^o$.
The deviations of the mixing angles from their tri-bimaximal values 
can be expressed as,
\begin{eqnarray}
&& \theta_{13} = \Delta^{TB}_{13}, \ \ 
\theta_{12} = 35^o+\Delta^{TB}_{12}, \ \  
\theta_{23}  =  45^o+\Delta^{TB}_{23}. \label{rsadegrees}
\end{eqnarray}

\section{Family Symmetry}

\begin{figure}[htb]
\centering
\includegraphics[width=0.46\textwidth]{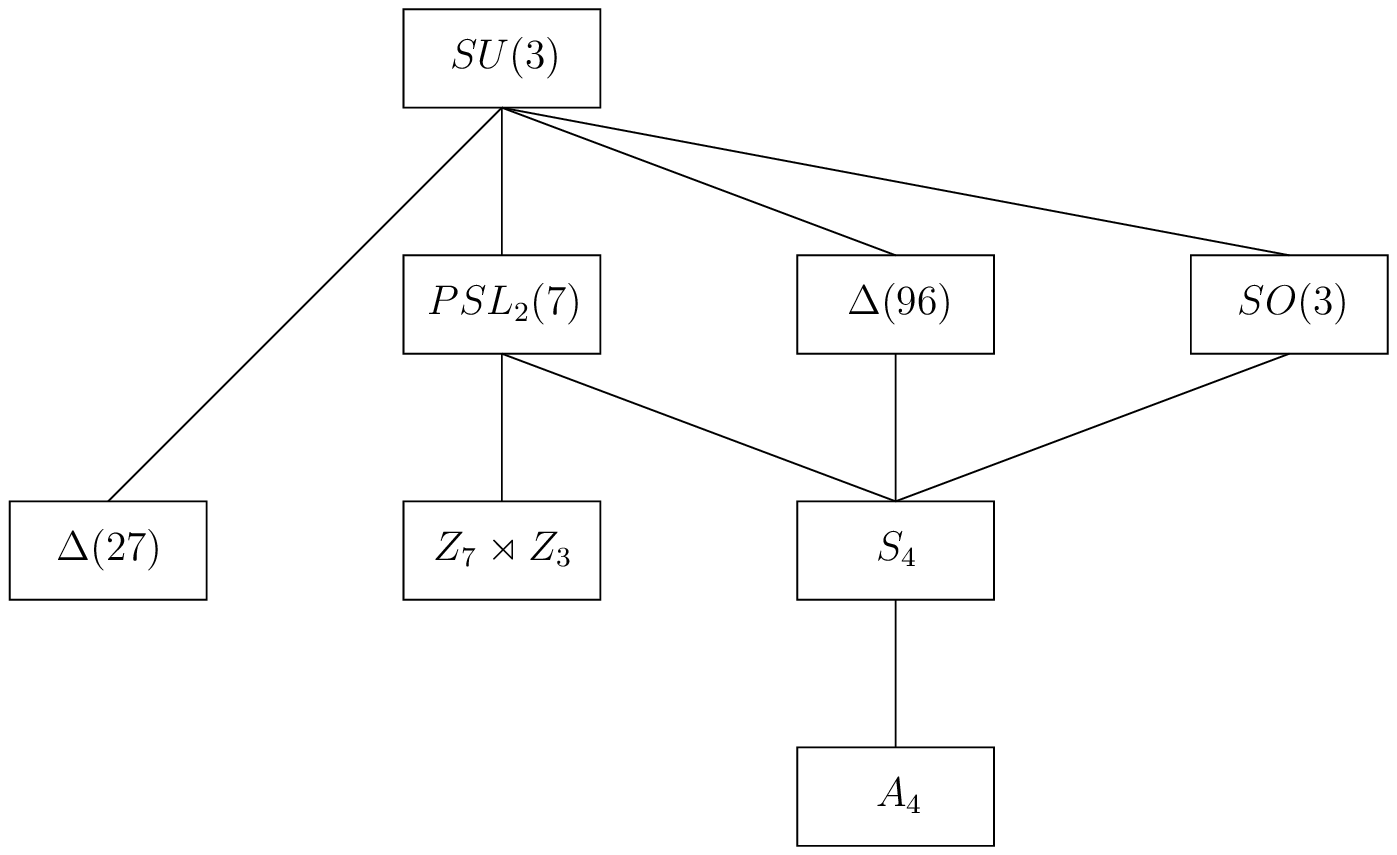}
\vspace*{-4mm}
    \caption{Some subgroups of $SU(3)$ that contain triplet representations and have been used
    as candidate family symmetries $G_f$. } \label{family}
\vspace*{-2mm}
\end{figure}

Let us expand the neutrino mass matrix in the diagonal charged
lepton basis, assuming exact TB mixing, as
${M^{\nu}_{TB}}=U_{TB}{\rm diag}(m_1, m_2, m_3)U_{TB}^T$ leading
to (absorbing the Majorana phases in $m_i$):
\begin{equation}
\label{mLL} {M^{\nu}_{TB}}= m_1\Phi_1 \Phi_1^T + m_2\Phi_2 \Phi_2^T + m_3\Phi_3 \Phi_3^T
\end{equation}
where $\Phi_1^T=\frac{1}{\sqrt{6}}(2,-1,1)$,
$\Phi_2^T=\frac{1}{\sqrt{3}}(1,1,-1)$, $\Phi_3^T=\frac{1}{\sqrt{2}}(0,1,1)$,
are the respective columns of $U_{TB}$
and $m_i$ are the physical neutrino masses. In the neutrino
flavour basis (i.e. diagonal charged lepton mass basis), it has
been shown that the above TB neutrino mass matrix is invariant
under $S,U$ transformations:
\begin{equation}
{M^{\nu}_{TB}}\,= S {M^{\nu}_{TB}} S^T\,= U {M^{\nu}_{TB}} U^T \ .
\label{S} \end{equation}
A very straightforward argument
\cite{King:2009ap}
shows that this neutrino flavour symmetry group
has only four elements corresponding to Klein's four-group $Z_2^S
\times Z_2^U$. By contrast the diagonal charged lepton mass matrix
(in this basis) satisfies a diagonal phase symmetry $T$. The
matrices $S,T,U$ form the generators of the group $S_4$ in the
triplet representation, while the $A_4$ subgroup is generated by
$S,T$. Some candidate family symmetries $G_f$ are shown in Fig.\ref{family}.

%%%%%%%%%%%%%%%%%%%%%%%%%%%%%%%%%%%%%%%%%%%%%%%%%%%%%%%%%%%%%%%%%%%%%%
%%%%%%%%%%%%%%%%%%%%%%%%%%%%%%%%%%%%%%%%%%%%%%%%%%%%%%%%%%%%%%%%%%%%%%

\section{Direct vs Indirect Models and Form Dominance}

As discussed in \cite{King:2009ap},
the flavour symmetry of the neutrino mass matrix may originate
from two quite distinct classes of models. The first class of models,
which we call direct models, are based on a family symmetry $G_f=S_4$,
or a closely related family symmetry as discussed below,
some of whose generators are directly preserved in the
lepton sector and are manifested as part of the observed flavour
symmetry. The second class of models, which we call indirect
models, are based on some more general family symmetry $G_f$ which is completely
broken in the neutrino sector, while the observed neutrino flavour
symmetry $Z_2^S \times Z_2^U$ in the neutrino flavour basis emerges as an
accidental symmetry which
is an indirect effect of the family symmetry $G_f$. In such
indirect models the flavons responsible for the neutrino masses break $G_f$
completely so that none of the generators of $G_f$ survive in the
observed flavour symmetry $Z_2^S \times Z_2^U$.

In the direct models, the symmetry of the neutrino mass matrix in
the neutrino flavour basis (henceforth called the neutrino mass matrix for
brevity) is a remnant of the $G_f=S_4$ symmetry of the Lagrangian,
where the generators
$S,U$ are preserved in the neutrino sector, while the diagonal generator
$T$ is preserved in the charged lepton sector.
For direct models, a larger family symmetry $G_f$ which
contains $S_4$ as a subgroup is also possible e.g. $G_f=PSL(2,7)$ \cite{King:2009mk}.
Typically direct models satisfy form
dominance \cite{Chen:2009um,Choubey:2010vs}, and require flavon F-term
vacuum alignment, permitting an $SU(5)$ type unification
\cite{Altarelli:2005yp} typically based in $A_4$ family symmetry \cite{Ma:2001dn}. 
Such minimal $A_4$ models lead to
neutrino mass sum rules between the three masses $m_i$, resulting
in/from a simplified mass matrix in Eq.\ref{mLL}. $A_4$ may result
from 6D orbifold models \cite{Altarelli:2006kg} and recently an
$A_4\times SU(5)$ SUSY GUT model has been constructed in 6D \cite{Burrows:2009pi},
while a similar model in 8D enables vacuum alignment to be elegantly 
achieved by boundary conditions \cite{Burrows:2010wz}.

In the indirect models \cite{King:2009ap} the idea is that the three columns of $U_{TB}$
$\Phi_i$ are promoted to new Higgs fields called
``flavons'' whose VEVs break the family symmetry, with the
particular vacuum alignments along the directions
$\Phi_i$. In the indirect models
the underlying family symmetry of the Lagrangian $G_f$ is
completely broken, and the flavour symmetry of the neutrino mass
matrix $Z_2^S \times Z_2^U$ emerges entirely as an accidental
symmetry, due to the presence of flavons with particular vacuum
alignments proportional to the columns of $U_{TB}$, where such
flavons only appear quadratically in effective Majorana
Lagrangian \cite{King:2009ap}. Such vacuum alignments can be
elegantly achieved using D-term vacuum alignment, which allows
the large classes of discrete family symmetry $G_f$,
namely the $\Delta(3n^2)$ and $\Delta(6n^2)$ groups \cite{King:2009ap}.
The indirect models satisfy natural form dominance since each column
of the Dirac mass matrix corresponds to a different flavon VEV.
In the limit $m_1\ll m_2 < m_3$ FD reduces to constrained sequential
dominance (CSD)\cite{King:2005bj}. Examples of discrete symmetries used in
the indirect approach can be found in \cite{deMedeirosVarzielas:2005qg}.

Explicitly, the TB form of the neutrino mass matrix
in Eq.\ref{mLL} is obtained from the see-saw mechanism in these models as follows. 
In the diagonal right-handed neutrino mass basis we
may write $M_{RR}^{\nu}={\rm diag}(M_A, M_B, M_C)$ and the Dirac
mass matrix as $M_{LR}^{\nu}=(A,B,C)$ where $A,B,C$ are three
column vectors. Then the type I see-saw formula
${M^{\nu}}=M_{LR}^{\nu}(M_{RR}^{\nu})^{-1}(M_{LR}^{\nu})^T$
gives
\begin{equation}
\label{mLLCSD} {M^{\nu}}=\frac{AA^T}{M_A}+ \frac{BB^T}{M_B} +
\frac{CC^T}{M_C}.
\end{equation}
By comparing Eq.\ref{mLLCSD} to the TB form in Eq.\ref{mLL} it is
clear that TB mixing will be achieved if $A\propto \Phi_3$,
$B\propto \Phi_2$, $C\propto \Phi_1$, with each of $m_{3,2,1}$
originating from a particular right-handed neutrino of mass
$M_{A,B,C}$, respectively. This mechanism
allows a completely general neutrino mass spectrum and, since the
resulting ${M^{\nu}}$ is form diagonalizable, it is referred to
as form dominance (FD) \cite{Chen:2009um}. For example, 
the direct $A_4$ see-saw models
\cite{Altarelli:2005yp} satisfy FD \cite{Chen:2009um}, where each
column corresponds to a linear combination of flavon VEVs.

A more natural possibility, called Natural FD, arises when each column
arises from a separate flavon VEV, and this possibility corresponds to
the case of indirect models. For example,
if $m_1\ll m_2 < m_3$ then the precise form of $C$ becomes
irrelevant, and in this case FD reduces to constrained sequential
dominance (CSD)\cite{King:2005bj}. The CSD mechanism has been
applied in this case to the class of indirect models
with Natural FD based on the family symmetries
$SO(3)$ \cite{King:2005bj,King:2006me} and $SU(3)$
\cite{deMedeirosVarzielas:2005ax}, and their discrete subgroups
\cite{deMedeirosVarzielas:2005qg}.

 \section{Tri-bimaximal mixing and GUTs}
Tri-bimaximal mixing (even if only approximately realised) 
seems to suggest an underlying non-Abelian discrete family symmetry such as $S_4$
that might unlock the long-standing flavour puzzle. However, 
when combined with Grand Unification such as $SU(5)$, the tri-bimaximal
mixing prediction is always violated due to the requirement of 
non-zero quark mixing 
\cite{King:2006np,deMedeirosVarzielas:2006fc,Altarelli:2008bg,Chen:2007afa,Hagedorn:2010th}. 
The resulting mixing from the charged lepton sector will always
lead to deviations from tri-bimaximal lepton mixing, for example
resulting in a non-zero reactor angle of about $\theta_{13}^o \approx 3^o$.
{\em Note that the reactor angle cannot be equal to zero in 
TBM $\otimes$ GUT models.} 

Moreover, TBM $\otimes$ GUT models
predict sum rule relations between the deviation parameters such as 
\cite{King:2005bj,Masina:2005hf,Antusch:2005kw,Boudjemaa:2008jf,Antusch:2007rk}:
\beq
\Delta_{12}^{TB}\approx \theta_{13} \cos \delta, \ \ \theta_{13}\approx \frac{\theta_C}{3\sqrt{2}}
\approx 3^o
\eeq
where $\theta_C=13^o$ is the Cabibbo angle and $\delta$ is the observable
CP violating oscillation phase, with RG corrections of less than
one degree. 
Such sum rules provide a motivation for 
future high precision neutrino oscillation experiments 
capable of measuring the reactor angle and the CP violating oscillation phase,
as well as the deviation of the solar angle from its tri-bimaximal value
\cite{Bandyopadhyay:2007kx}.

In certain classes of TBM $\otimes$ GUT models, the leptonic CP violating phase $\delta$
is related to the CKM unitarity triangle angle $\alpha \approx 90^o$.
In practice the Family Symmetry $\otimes$ GUT models which predict 
$\alpha = 90^o$ also predict a discrete set of possibilities for the CP violating oscillation phase
 $\delta = 0^o, 90^o, 180^o, 270^o$ \cite{Antusch:2011sx}.
These discrete possibilities could be distinguished by future high precision neutrino oscillation experiments.

\section{Tri-bimaximal-reactor Mixing}
If the reactor angle is measured to be large, but the solar and atmospheric angles remain close to their tri-bimaximal
values, i.e. the deviation parameters in Eq.\ref{rsa} take the form
$s=a=0$ but $r\neq 0$, then the mixing matrix takes the ``tri-bimaximal-reactor'' (TBR) form
\cite{King:2009qt}:
\begin{eqnarray}
U_{TBR} =
\left( \begin{array}{ccc}
\sqrt{\frac{2}{3}}  & \frac{1}{\sqrt{3}} & \frac{1}{\sqrt{2}}re^{-i\delta } \\
-\frac{1}{\sqrt{6}}(1+ re^{i\delta })  & \frac{1}{\sqrt{3}}(1- \frac{1}{2}re^{i\delta })
& \frac{1}{\sqrt{2}} \\
\frac{1}{\sqrt{6}}(1- re^{i\delta })  & -\frac{1}{\sqrt{3}}(1+ \frac{1}{2}re^{i\delta })
 & \frac{1}{\sqrt{2}}
\end{array}
\right)P.
\label{MNS3}
\end{eqnarray}
Such TBR pattern may be accomplished via partially constrained sequential dominance
(PCSD) as a variation of Eq.\ref{mLLCSD} in the hierarchical limit where $m_1\ll m_2$ where 
$C$ is irrelevant and by taking 
$B^T=\frac{1}{\sqrt{3}}(1,1,-1)$, $A^T=\frac{1}{\sqrt{2}}(\varepsilon,1,1)$,
where $\varepsilon$ is a small correction to the vacuum alignment, leading to 
$\varepsilon = re^{-i\delta }$ \cite{King:2009qt}.

\section{Quark-Lepton Complementarity}
If the reactor angle is measured to be larger than the above prediction and the solar
and atmospheric angles deviate significantly from their TB values, then 
one may consider as a starting point bimaximal mixing which corresponds to $ \theta_{12} = 45^o$, 
$\theta_{23}  =  45^o$ and $\theta_{13}  =  0^o$.
Non-Abelian discrete family symmetry models based on $S_4$
can alternatively lead to bimaximal mixing if the requirement of GUTs is relaxed \cite{Altarelli:2009gn}. 
Such models require large deviations and these large deviations imply a large reactor angle, typically  
$\theta_{13} \approx \theta_C \approx 13^o$.

The deviations of the mixing angles from their bimaximal values 
can be expressed as,
\begin{eqnarray}
&& \theta_{12} = 45^o+\Delta^{BM}_{12}, \ \  
\theta_{23}  =  45^o+\Delta^{BM}_{23}. \label{rsadegrees}
\end{eqnarray}
The experimentally measured solar angle requires a large deviation
corresponding to the sum rule relation \cite{Antusch:2005kw}, 
\beq
\Delta_{12}^{BM}\approx \theta_{13} \cos \delta, \ \ \theta_{13}\approx \theta_C.
\eeq
Assuming $\delta \approx 180^o$,
this would imply,
\beq
 \theta_{12} +  \theta_C = 45^o,
\eeq
known as ``quark-lepton complementarity'' (QLC).
There is no straightforward GUT model that can achieve QLC.

\section{Abelian Family Symmetry}
\begin{figure}[h]
\begin{minipage}{18pc}
\includegraphics[width=18pc]{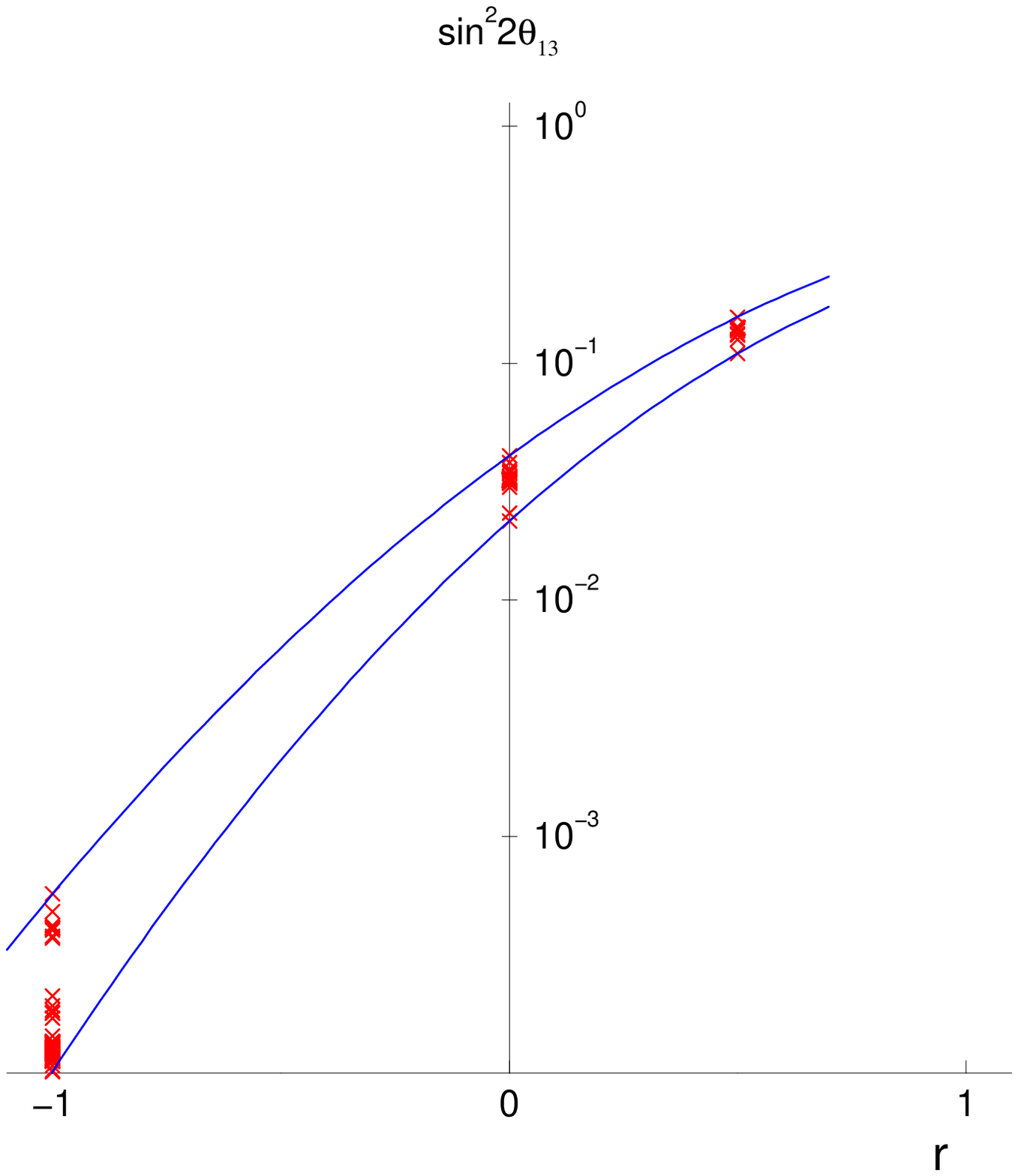}
\caption{\label{blazek}
The prediction for $\sin^22\theta_{13}$ in Abelian family symmetry models (normal hierarchy)
as a function of an undetermined ratio of Yukawa couplings $r$.  }
  \end{minipage} 
   \hspace{2pc}
  \begin{minipage}{18pc}
\includegraphics[width=18pc]{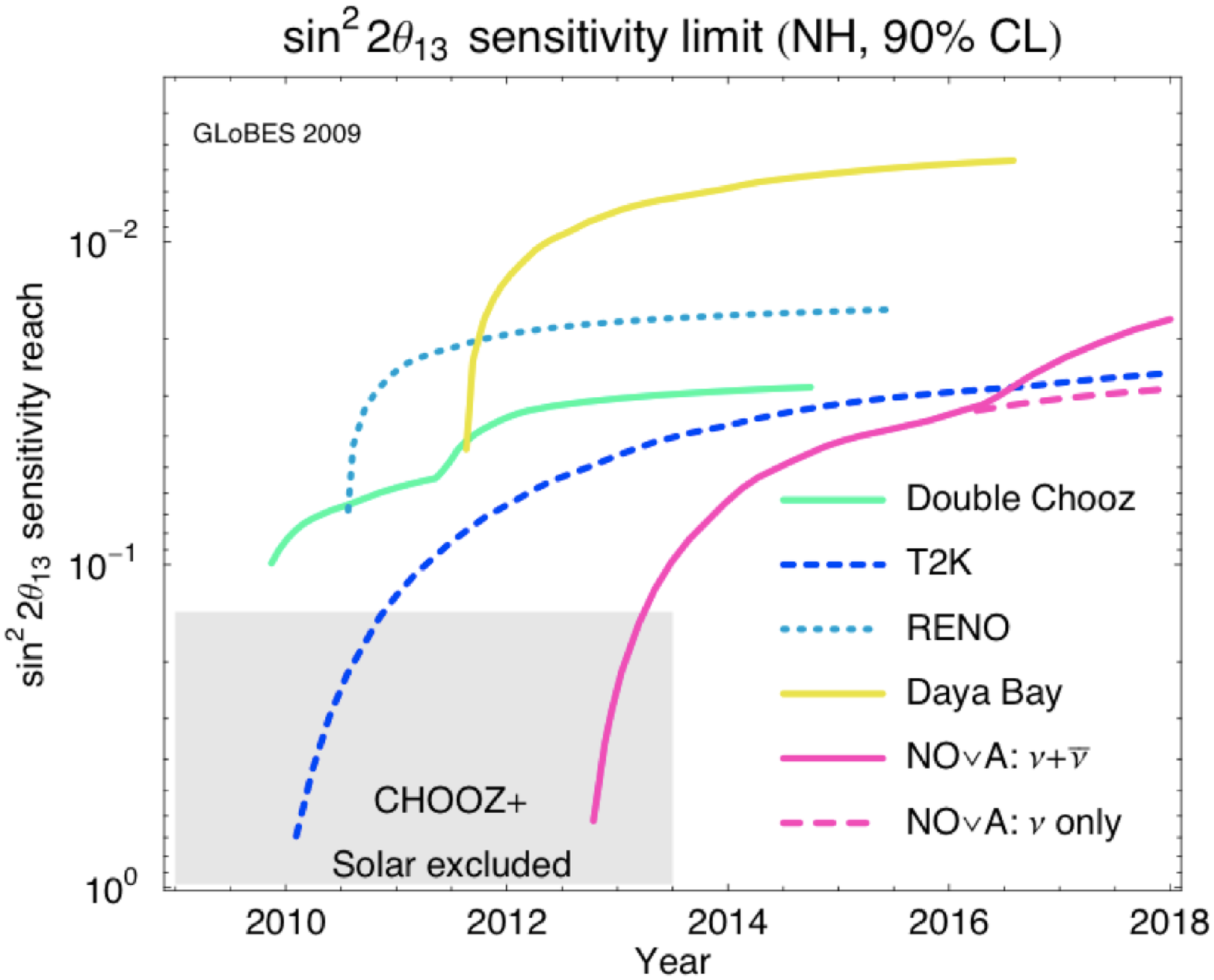}
\caption{\label{future} The future $\sin^22\theta_{13}$ sensitivity limit (normal hierarchy, 90\% CL). }
\end{minipage}%
\end{figure}

With a simple $U(1)$ family symmetry it is impossible to obtain exact
tri-bimaximal or bimaximal mixing, except by accident.
On the other hand it is easy to explain qualitative features such as,
\begin{eqnarray}
&& \theta_{12} \sim  {\rm large}, \ \  
\theta_{23}  \sim  {\rm large},  \ \ 
\theta_{13} \sim  {\rm small}.\label{rough}
\end{eqnarray}
For example, for a hierarchical spectrum the typical expectation of see-saw models with 
sequential dominance models approximately given by \cite{Blazek:2002wq}
\beq
\theta_{13} \sim  O\left( \frac{m_2}{m_3}\right). 
\eeq
However the precise prediction depends on an undetermined ratio of Yukawa couplings $r$ as shown in 
Fig.\ref{blazek}, which may be compared to the future experimental sensitivities in Fig.\ref{future}
\cite{Huber:2009cw}.

%\begin{figure}
%\vspace*{13pt}
%\leftline{\hfill\vbox{\hrule width 5cm height0.001pt}\hfill}
   %   \mbox{\epsfig{figure=blazek.eps,width=6.0cm}}
%\vspace*{1.4truein}		%ORIGINAL SIZE=1.6TRUEIN x 100% - 0.2TRUEIN
%\leftline{\hfill\vbox{\hrule width 5cm height0.001pt}\hfill}
%\caption{The prediction for $\sin^22\theta_{13}$ in Abelian family symmetry models (normal hierarchy)
%as a function of an undetermined ratio of Yukawa couplings $r$. 
%from \cite{Blazek:2002wq}.
%}
%\label{blazek}
%\end{figure}

\section{Summary}

Over the past dozen years there has been a revolution in our understanding of neutrino physics.
Yet, despite this progress, it must be admitted that we still do not understand the origin 
or nature of neutrino mass and mixing. 
However it is a striking fact that current data on lepton mixing is
consistent with the so-called tri-bimaximal mixing pattern and many models have been proposed.
Realistic models predict various deviations from
tri-bimaximal mixing, for example the benchmark models shown in Table \ref{mod}.

\begin{table}[htdp]
\begin{center}
\begin{tabular}{|c|c|c|c|c|}
\hline
Benchmark Model & $\theta_{13}$ & $|\theta_{23}-45^o|$  & $|\theta_{12}-35^o|$ & $\delta$  \\
\hline
TBM $\otimes$ GUT \cite{Antusch:2011sx} & $\frac{\theta_C}{3\sqrt{2}}=3^o$ 
& $\leq 1^o$ &  $\leq 1^o$ & $90^o, 270^o$  \\
\hline
TBR \cite{King:2009qt} & any &  $\leq 1^o$ &  $\leq 1^o$ & any \\
\hline
QLC \cite{Altarelli:2009gn}  & $\theta_C =13^o$ & $\leq 1^o$ &  large & $180^o$  \\
\hline
Abelian \cite{Blazek:2002wq} & Fig.\ref{blazek} & large & large & any \\
\hline
\end{tabular}
\end{center}
\caption{Predictions of benchmark models for the deviations from tri-bimaximal mixing and $\delta$.}
\label{mod}
\end{table}%

In order to discriminate between the benchmark models in Table \ref{mod},
and hence shed light on GUT models of Flavour,
it is necessary to measure
the deviations of the reactor, solar and atmospheric angles from 
their tri-bimaximal values, as well as $\delta$.
From a theorists' perspective the job is not done until the deviations from tri-bimaximal mixing 
and $\delta$ are measured.
This will require high precision neutrino oscillation experiments,
based on a next generation neutrino accelerator \cite{Bandyopadhyay:2007kx}.

Postscript: while writing these proceedings T2K have published evidence for a 
large non-zero reactor angle \cite{:2011sj}. If confirmed these results would have major implications
for neutrino mass models as discussed above.

\section{Acknowledgements}
I would like to thank the local organisers of Neutel 2011 in Venice,
in particular Mauro Mezzetto for a very enjoyable and stimulating conference.

 \end{document}